\documentclass[pra,twocolumn,aps,amssymb,showpacs,superscriptaddress]{revtex4-1}

\usepackage{epsfig}
\usepackage{graphicx}
\usepackage{amsmath} 
\usepackage{amssymb}
\usepackage{enumerate}
\usepackage{hyperref}
\usepackage[normalem]{ulem}
\usepackage{cancel}

\usepackage{color}
\definecolor{rosso}{rgb}{1,0,0}
\definecolor{verde}{rgb}{0,1,0}
\definecolor{blue}{rgb}{0,0,1}
\definecolor{verdescuro}{rgb}{0,0.5,0.5}
\definecolor{rossoscuro}{rgb}{0.7,0.3,0}
\definecolor{bluscuro}{rgb}{0.3,0,0.7}
\definecolor{magenta}{rgb}{1,0,1}

\begin{document}

\title{Inclusion of pairing fluctuations in a semiclassical approach: \\ The case of study of the Josephson effect}

\author{V. Piselli}
\affiliation{School of Science and Technology, Physics Division, Universit\`{a} di Camerino, 62032 Camerino (MC), Italy}
\affiliation{CNR-INO, Istituto Nazionale di Ottica, Sede di Firenze, 50125 (FI), Italy}
\author{L. Pisani}
\affiliation{School of Science and Technology, Physics Division, Universit\`{a} di Camerino, 62032 Camerino (MC), Italy}
\author{G. Calvanese Strinati}
\email{giancarlo.strinati@unicam.it}
\affiliation{School of Science and Technology, Physics Division, Universit\`{a} di Camerino, 62032 Camerino (MC), Italy}
\affiliation{CNR-INO, Istituto Nazionale di Ottica, Sede di Firenze, 50125 (FI), Italy}

\begin{abstract}
Recent refinements on a semiclassical approach are reviewed, aiming at describing the inhomogeneous local gap parameter in the presence of non-trivial spatial geometries and at taking into account at the same time pairing fluctuations beyond mean field. 
The method is applied to describe the Josephson effect over the wide range of physical conditions related to recent experiments on this topic performed with ultra-cold Fermi gases.
\end{abstract}

\maketitle

\section{Introduction} 
\label{sec:introduction}

Professor Peter Schuck has long been interested in deriving from first principles semiclassical differential equations for the inhomogeneous local gap parameter $\Delta(\mathbf{r})$ in terms of an expansion in the reduced Planck constant $\hbar$ 
of the generalized density matrix for superfluid inhomogeneous Fermi systems.
His first pioneering article along these lines goes back to 1992 \cite{Taruishi-1992}, and his interest on this topic continued until the very last part of his life \cite{Schuck-2023}.
In particular, in Ref.~\cite{Pei-2015} numerical calculations were implemented by Peter Schuck and co-workers which showed that the coarse-graining LPDA approach of Ref.~\cite{Simonucci-2014}
(with the acronym LPDA standing for Local Phase Density Approximation),
where only a few terms of the expansion were effectively retained, could provide reasonable results when compared with more complete approaches.
This was probably the motivation for Schuck's late interest in the approach of Ref.~\cite{Simonucci-2014}, about which the senior author (G.~C.~S.) of the present article has had frequent and detailed discussions with Peter Schuck especially during the year 2021, discussions which unfortunately had to stop abruptly owing to Schuck's sudden health problems in the Spring 2022.

Specifically, the senior author of the present article regrets that he did not have the opportunity for updating Peter Schuck about the progresses that were just beginning to develop at the time of their last meeting,
concerning the inclusion of pairing fluctuation corrections on top of the mean-field LPDA approach of Ref.~\cite{Simonucci-2014}.
Here, we aim at summarizing those progresses that have materialized in the meantime and present their recent numerical implementation in the context of the Josephson effect.
A more extensive account of these results can be found in Refs.~\cite{Pisani-2023,Piselli-2023}.

Before deepening in these late results, it is first worth recalling the motivations behind the semiclassical approach by Peter Schuck \cite{Taruishi-1992,Pei-2015,Schuck-2023} and the original LPDA approach \cite{Simonucci-2014}
to the Bogoliubov-deGennes (BdG) equations \cite{BdG-1966}, which were both meant to replace the BdG equations by a second order differential equation directly for the local gap parameter $\Delta(\mathbf{r})$.
The main reason behind this replacement is the large computation time and huge memory space needed when solving numerically the BdG equations, especially in the presence of non-trivial geometrical constraints and when spanning the BCS-BEC crossover, from the Bardeen-Cooper-Schrieffer (BCS) limit of highly overlapping Cooper pairs to the Bose-Einstein condensate (BEC) limit of dilute composite bosons \cite{Phys-Rep-2018}.
In particular, the LPDA equation obtained in Ref.~\cite{Simonucci-2014} through a suitable double coarse-graining procedure on BdG equations has proved capable of reducing the computation time to even $10^{5}$ times shorter 
than that required when solving BdG equations.
In addition, the LPDA equation recovers the Ginzburg-Landau equation in the BCS limit at high temperature and the Gross-Pitaevskii equation in the BEC limit at zero temperature.

Notwithstanding these important achievements, the LPDA approach, being based on an approximation to the BdG equations which are themselves based on a mean-field decoupling of the Hamiltonian, misses the beyond-mean-field effects of pairing
fluctuations which are known to be relevant when departing from the BCS limit \cite{Phys-Rep-2018}.
In addition, what complicates the picture is that to go beyond the LPDA approach, these beyond-mean-field effects of pairing fluctuations have to be included over and above an inhomogeneous geometrical environment.
This problem was recently addressed in Ref.~\cite{Pisani-2023}, where a \emph{modified\/} mLPDA approach was proposed (with m standing for modified) to implement the inclusion of pairing fluctuations in the LPDA approach.
In this context, pairing fluctuations were mainly considered at the level of the non self-consistent $t$-matrix approximation in the superfluid phase developed in Ref.~\cite{PPS-2004}, although improvements \cite{PPSb-2018} on that approximation
were also considered in Ref.~\cite{Pisani-2023}.
At the same time, in the related Ref.~\cite{Piselli-2023} the mLPDA approach was implemented to deal specifically with the Josephson effect recently studied experimentally in ultra-cold Fermi gases in Refs.~\cite{Kwon-2020,DelPace-2021}.
In the present article, we will first briefly summarize the essential features of the LPDA \cite{Simonucci-2014} and mLPDA \cite{Pisani-2023} approaches for the reader's convenience, and then we will concentrate on the specific application of these 
approaches to the Josephson effect following Ref.~\cite{Piselli-2023}.
In particular, with the theoretical tools at our disposal, we will delve into the potential relation between the Josephson effect and the superfluid density, thereby contributing to fill an apparent gap in the current literature \cite{Modugno-2023}.

The present article is organized as follows. 
Section~\ref{sec:theoretical_approach} briefly recalls the LPDA and mLPDA approaches, having in mind their application to the Josephson effect where a supercurrent flowing through non-trivial geometrical constraints (typically, a barrier).
In this context, numerical results for the ensuing critical current are presented in Sec.~\ref{sec:Numerical results} when the barrier has a rather simple form, and in Sec.~\ref{sec:experiments} when the barrier has the complex shape utilized 
in the experiments with ultra-cold Fermi gases.
Section~\ref{sec:Josephson_vs_superf-dens} delves specifically into the potential relation between the Josephson effect and the superfluid density.
Section~\ref{sec:conclusions} gives our conclusions.
Throughout we shall consider balanced spin populations and set $\hbar=1$ for simplicity.

\section{Theoretical approach} 
\label{sec:theoretical_approach}

In this Section, we briefly recall the LPDA approach introduced in Ref.~\cite{Simonucci-2014} at the mean-field level from a (double) coarse graining of the BdG equations, 
and its refined mLPDA approach introduced in Ref.~\cite{Pisani-2023} where pairing fluctuations are also included at a local level, so as to deal with inhomogeneous spatial environments and beyond-mean-field effects on equal footing.
Both approaches will be specified in the presence of a superfluid flow for applications to the Josephson effect.

In the following, the BCS-BEC crossover will be spanned in terms of the dimensionless parameter $(k_Fa_F)^{-1}$, where $a_F$ is the scattering length for the two-fermion problem in vacuum and $k_F=(3\pi^2n)^{1/3}$ is the Fermi wave vector with (uniform) particle density $n$. 
This parameter ranges from $(k_Fa_F)^{-1}\leq-1$ in the weak-coupling (BCS) regime when $a_F<0$ to $(k_Fa_F)^{-1}\geq+1$ in the strong-coupling (BEC) regime when $a_F>0$ across the unitary limit when $|a_F|$ diverges \cite{Phys-Rep-2018}.

\vspace{0.05cm}
\begin{center}
{\bf A. The LPDA equation for the Josephson effect \\ with a simple slab geometry}
\end{center}

We initially consider the Josephson effect with a slab geometry, where a supercurrent flows through a barrier of limited spatial width in the flow direction (say, $x$) and is homogeneous in the orthogonal $y$ and $z$ directions.
In this case, it is convenient to write the local gap parameter $\Delta(x)$ in the form \cite{PSS-2020}
\begin{equation}
\Delta(x)=|\Delta(x)|e^{2i\mathbf{q}\cdot\mathbf{x}+2i\phi(x)}=e^{2i\mathbf{q}\cdot\mathbf{x}}\tilde{\Delta}(x) \, ,
\label{eq:delta_LPDA}
\end{equation}
\noindent
where $\mathbf{q} = q \hat{x}$.
Note that the phase of the gap parameter is the sum of two terms, namely, $2\mathbf{q}\cdot\mathbf{x}$ associated with the superfluid flow even in the absence of the barrier \cite{BdG-1966}
and $2\phi(x)$ due to the presence of the barrier. 
Under these circumstances, the LPDA equation takes the form \cite{PSS-2020}
\begin{equation}
\begin{split}
-\dfrac{m}{4\pi a_F}\tilde\Delta(x)&=\mathcal{I}_0(\mathbf{x})\tilde\Delta(x)+\dfrac{\mathcal{I}_1(x)}{4m}\dfrac{\text{d}^2}{\text{d}x^2}\tilde\Delta(x)\\
&+i\mathcal{I}_1(x)\dfrac{q}{m}\dfrac{\text{d}\tilde{\Delta}(x)}{\text{d}x} 
\label{eq:LPDA}
\end{split}
\end{equation}
\noindent
with the coefficients $\mathcal{I}_0$ and $\mathcal{I}_1$ given by
\begin{subequations}\label{eq:I_0I_1}
\begin{equation}\label{eq:I_0}
\mathcal{I}_0(x)=\int\!\!\dfrac{\text{d}\mathbf{k}}{(2\pi)^3}\left[\dfrac{1-2f_F(E_+^\mathbf{Q_0}(\mathbf{k}|x))}{2E(\mathbf{k}|x)}-\dfrac{m}{\mathbf{k}^2}\right]
\end{equation}
\begin{equation}\label{eq:I_1}
\begin{split}
\mathcal{I}_1(x)&=\dfrac{1}{2}\int\!\!\dfrac{\text{d}\mathbf{k}}{(2\pi)^3}\bigg\{\dfrac{\xi(\mathbf{k}|x)}{2E(\mathbf{k}|x)^3}[1-2f_F(E_+^\mathbf{q}(\mathbf{k}|x))]\\
&+\dfrac{\xi(\mathbf{k}|x)}{2E(\mathbf{k}|x)^2}\dfrac{\partial f_F(E_+^\mathbf{Q_0}(\mathbf{k}|x))}{\partial E_+^\mathbf{q}(\mathbf{k}|x)}\\
&+\dfrac{\mathbf{k}\cdot\mathbf{q}}{\mathbf{q}^2}\dfrac{1}{E(\mathbf{k}|x)}\dfrac{\partial f_F(E_+^\mathbf{Q_0}(\mathbf{k}|x))}{\partial E_+^\mathbf{q}(\mathbf{k}|x)}\bigg\}
\end{split}
\end{equation}
\end{subequations}
\noindent
where
\begin{equation}
\begin{split}
\xi(\mathbf{k}|x)&=\dfrac{\mathbf{k}^2}{2m}-\left[\mu -V_{\mathrm{ext}}(x)-\dfrac{\mathbf{q}^2}{2m}\right],\\
E(\mathbf{k}|x)&=\sqrt{\xi(\mathbf{k}|x)^2+|\tilde{\Delta}(x)|^2},\\ E_+^\mathbf{q}(\mathbf{k}|x)&=E(\mathbf{k}|x)+\dfrac{\mathbf{k}\cdot\mathbf{q}}{m} \, .
\label{eq:xiEE+}
\end{split}
\end{equation}

In addition, in Ref.~\cite{PSS-2020} the imaginary part of Eq.~\eqref{eq:LPDA} was replaced by the constraint of current conservation along the $x$ axis, which guarantees the validity of the continuity equation. 
In this approach, the current density takes the form
\begin{equation}
\begin{split}
j(x)&=\frac{1}{m}\left(\dfrac{\text{d}\phi(x)}{\text{d}x}+q \right)n(x)\\
&+2\int\!\!\dfrac{d\mathbf{k}}{(2\pi)^3}\dfrac{k_{x}}{m}f_{F}\left( E^{\mathbf{q}}_{+}(\mathbf{k}|x)\right)
\label{eq:current_LPDA}
\end{split}
\end{equation}
\noindent
where
\begin{equation}\label{eq:density_LPDA}
n(x)=\int\scalebox{0.9}[1]{$\!\!\dfrac{d\mathbf{k}}{(2\pi)^3}\left\{1-\dfrac{\xi^{\mathbf{q}}(\mathbf{k}|x)}{E^{\mathbf{q}}(\mathbf{k}|x)}\left[1-2 f_F(E^{\mathbf{q}}_+(\mathbf{k}|x))\right]\right\} $}
\end{equation}
is the corresponding local number density. 
In the expressions \eqref{eq:xiEE+} entering Eqs.~\eqref{eq:current_LPDA} and \eqref{eq:density_LPDA} $q$ is replaced by $q+\text{d}\phi/\text{d}x$.

\vspace{0.05cm}
\begin{center}
{\bf B. The mLPDA approach for the Josephson \\ effect with a simple slab geometry}
\end{center}

It is known that a consistent description of the BCS–BEC crossover cannot proceed without a proper inclusion of pairing fluctuations over and above mean field.
This is because, when approaching the BEC limit where truly bosonic pairs are formed out of opposite-spin fermions, while the mean field can describe only the ``internal'' degrees of freedom of these bosons, pairing fluctuations are required to account for their ``translational'' (center-of-mass) degrees of freedom and thus for their dynamics \cite{Phys-Rep-2018}.
This feature was originally pointed out by Nozi\`eres and Schmitt-Rink (NSR), as being required for obtaining the expected value of the Bose-Einstein critical temperature in the BEC limit of the BCS-BEC crossover \cite{NSR-1985}.
The NSR approach has since been much extended in the literature, by relying on the $t$-matrix approximation which was especially considered for the normal phase above the superfluid critical temperature \cite{Phys-Rep-2018}.
An extension to the super-fluid phase was also developed, in particular in terms of the non self-consistent t-matrix approximation \cite{PPS-2004}.
In all these approaches, however, homogeneous situations were only considered.
In this context, the mLPDA approach developed in Ref.~\cite{Pisani-2023} aims at including pairing fluctuations in the super-fluid phase over and above an \emph{inhomogeneous\/} mean field, 
by combining the (mean-filed) LPDA approach of Ref.~\cite{Simonucci-2014} with the treatment of pairing fluctuations within the non self-consistent t-matrix approximation of Ref.~\cite{PPS-2004}.

Accordingly, we recall that, when dealing with homogeneous Fermi superfluids in the absence of a superfluid flow, in Ref.~\cite{PPS-2004} the gap and density equations were treated on a different footing, namely, 
the gap equation was retained in its form at the mean-field level while the density equation was modified by including pairing fluctuations. 
By a similar token, for inhomogeneous Fermi superfluids but now also \emph{in the presence of a superfluid flow\/}, we here retain the LPDA equation for the superfluid gap in the form \eqref{eq:LPDA} 
and include pairing fluctuations in the expressions of both the local particle and current densities. 
This is the way how the mLPDA approach was set up in Ref.~\cite{Pisani-2023} starting from the LPDA approach of Ref.~\cite{Simonucci-2014}.

Quite generally, the required expressions of the particle and current densities can be written as follows in terms of the diagonal (normal) single-particle Green's function:
\begin{subequations}
\begin{equation}
n(x)=\dfrac{2}{\beta}\sum_ne^{i\omega_n\eta} G_{11}(x,x';\omega_n)
\label{eq:density_Green}
\end{equation}
\begin{equation}
j(x)=\dfrac{1}{\beta}\sum_n\dfrac{e^{i\omega_n\eta}}{im}\left(\dfrac{\text{d}}{\text{d}x}-\dfrac{\text{d}}{\text{d}x'}\right) G_{11}(x,x';\omega_n) 
\label{eq:current_Green}
\end{equation}
\end{subequations}
where the single-particle Green's function depends also on the wave vector $\mathbf{q}$ of Eq.~(\ref{eq:delta_LPDA}) in the presence of a supercurrent.
\noindent
Here, $\beta=(k_BT)^{-1}$ is the inverse temperature ($k_B$ being the Boltzmann constant), $\eta$ a positive infinitesimal, and $\omega_n=(2n+1)\pi/\beta$ ($n$ integer) a fermionic Matsubara frequency \cite{Schrieffer-1964}.
To comply with our needs, the Green's function in Eqs.~(\ref{eq:density_Green}) and (\ref{eq:current_Green}) has to take into account beyond-mean-field pairing fluctuations in the presence of a superfluid flow. 

As a first step, we consider the case of a homogeneous superfluid, for which in the presence of a supercurrent $G_{11}(x,x';\mathbf{q}) = e^{i \mathbf{q} \cdot (\mathbf{r} - \mathbf{r'})} \, \mathcal{G}_{11}(x-x';\mathbf{q})$ in 
Eqs.~(\ref{eq:density_Green}) and (\ref{eq:current_Green}).
In this case, to account for pairing fluctuations within the $t$-matrix approximation in the presence of a superfluid flow, one considers the reduced self-energies $\mathfrak{S}_{ii'}^{\mathrm{pf}}(k;\mathbf{q})$ \cite{Pisani-2023}
\begin{eqnarray}
\mathfrak{S}_{11}^{\mathrm{pf}}(k;\mathbf{q}) & = & -\mathfrak{S}_{22}^{\mathrm{pf}}(-k;\mathbf{q})=-\sum_Q\Gamma_{11}(Q;\mathbf{q}) \mathcal{G}_{11}^{\mathrm{mf}}(Q-k;\mathbf{q}) 
\nonumber \\
\mathfrak{S}_{12}^{\mathrm{pf}}(k;\mathbf{q}) & = & \mathfrak{S}_{21}^{\mathrm{pf}}(k;\mathbf{q})=-\Delta_{\mathbf{q}} \, ,
\label{eq:reduced}
\end{eqnarray}
\noindent
where $Q=(\mathbf{Q},\Omega_\nu)$ is a four-vector with $\Omega_\nu=2\nu\pi/\beta$ ($\nu$ integer) a bosonic Matsubara frequency,
\begin{equation*}
\sum_Q\longleftrightarrow\int\!\!\dfrac{\text{d}\mathbf{Q}}{(2\pi)^3}\dfrac{1}{\beta}\sum_\nu
\end{equation*}
is a short-hand notation, and the diagonal and off-diagonal single-particle Green's function at the mean-field (mf) level are given by:
\begin{subequations}\label{eq:Green_mf}
\begin{equation}
\mathcal{G}_{11}^{\mathrm{mf}}(k;\mathbf{q})=\dfrac{u(\mathbf{k};\mathbf{q})^2}{i\omega_n-E_+(\mathbf{k};\mathbf{q})}+\dfrac{v(\mathbf{k};\mathbf{q})^2}{i\omega_n+E_-(\mathbf{k};\mathbf{q})}
\end{equation}
\begin{equation}
\begin{split}
\mathcal{G}_{12}^{\mathrm{mf}}(k;\mathbf{q})&=-u(k;\mathbf{q})v(k;\mathbf{q})\left[\dfrac{1}{i\omega_n-E_+(\mathbf{k};\mathbf{q})}\right. \\
&\left.-\dfrac{1}{i\omega_n+E_-(\mathbf{k};\mathbf{q})}\right] \, .
\end{split}
\end{equation}
\end{subequations}
\noindent
\vspace{0.2cm}
Here,
\begin{subequations}
\begin{equation}
u(\mathbf{k};\mathbf{q})^2=\dfrac{1}{2}\left(1+\dfrac{\xi(\mathbf{k};\mathbf{q})}{E(\mathbf{k};\mathbf{q})}\right)
\end{equation}
\begin{equation}
v(\mathbf{k};\mathbf{q})^2=\dfrac{1}{2}\left(1-\dfrac{\xi(\mathbf{k};\mathbf{q})}{E(\mathbf{k};\mathbf{q})}\right)
\end{equation}
\end{subequations}
with the notation
\begin{equation}
\begin{split}
\xi(\mathbf{k};\mathbf{q})&=\dfrac{\mathbf{k}^2}{2m}-\mu+\dfrac{\mathbf{q}^2}{2m}\\
E(\mathbf{k};\mathbf{q})&=\sqrt{\xi(\mathbf{k};\mathbf{q})^2+\Delta_{\mathbf{q}}^2}\\
E_{\pm}(\mathbf{k};\mathbf{q})&=E(\mathbf{k};\mathbf{q})\pm\dfrac{\mathbf{k}\cdot\mathbf{q}}{m} \, .
\end{split}
\end{equation}
In addition, the elements of the particle-particle ladder $\Gamma_{ii'}(Q;\mathbf{q})$ in the broken-symmetry phase are given by \cite{Pisani-2023}
\begin{equation}
\scalebox{0.9}[1]{$
\begin{split}
\begin{bmatrix}
\Gamma_{11}(Q;\mathbf{q}) & \Gamma_{12}(Q;\mathbf{q}) \\
\Gamma_{21}(Q;\mathbf{q}) & \Gamma_{22}(Q;\mathbf{q})
\end{bmatrix}
&=\dfrac{1}{A(Q;\mathbf{q})A(-Q;\mathbf{q})-B(Q;\mathbf{q})^2}\\
&\times
\begin{bmatrix}
A(-Q;\mathbf{q}) & B(Q;\mathbf{q}) \\
B(Q;\mathbf{q}) & A(Q;\mathbf{q})
\end{bmatrix}
\end{split}
$}
\end{equation}
where
\begin{subequations}
\begin{equation}
\begin{split}
\scalebox{0.95}[1]A(Q;\mathbf{q})&=-\dfrac{m}{4\pi a_F}+\int\!\!\dfrac{d\mathbf{k}}{(2\pi)^3}\dfrac{m}{\mathbf{k}^2}\\
&-\sum_k\mathcal{G}_{11}^{\mathrm{mf}}(k+Q;\mathbf{q})\mathcal{G}_{11}^{\mathrm{mf}}(-k;\mathbf{q})
\end{split}
\end{equation}
\begin{equation}
B(Q;\mathbf{q})=\sum_k\mathcal{G}_{12}^{\mathrm{mf}}(k+Q;\mathbf{q})\mathcal{G}_{12}^{\mathrm{mf}}(-k;\mathbf{q}) \, .
\end{equation}
\end{subequations}
\noindent
Entering the expressions for the reduced self-energies \eqref{eq:reduced} into the Dyson's equation, the diagonal (normal) single-particle Green's function in the presence of a superfluid flow takes eventually the form:
\begin{widetext}
\vspace{0.3cm}
\begin{equation}
\mathcal{G}_{11}^{\mathrm{pf}}(k;\mathbf{q})=\dfrac{1}{i\omega_n-\xi(\mathbf{k}+\mathbf{q})-\mathfrak{S}_{11}^{\mathrm{pf}}(k;\mathbf{q})-\dfrac{\Delta_{\mathbf{q}}^2}{i\omega_n+\xi(\mathbf{k}-\mathbf{q})+\mathfrak{S}_{11}^{\mathrm{pf}}(-k;\mathbf{q})}} \, .
\label{eq:Green_pf}
\end{equation}
\end{widetext}

When considering the case of interest here of an inhomogeneous superfluid, in analogy to what was done in Ref.~\cite{Simonucci-2014} in the context of the LPDA approach, one may in principle obtain $G_{11}(x,x';\omega_n)$ needed in the expressions (\ref{eq:density_Green}) and (\ref{eq:current_Green}) by performing in Eq.~(\ref{eq:Green_pf}) the local replacements
\vspace{-0.3cm}
\begin{subequations}
\begin{equation}
\mu\longrightarrow\mu - V_\mathrm{{ext}}(x)
\label{eq:locmu}
\end{equation}
\begin{equation}
\Delta_{\mathbf{q}}\longrightarrow|\Delta(x)|
\end{equation}
\begin{equation}
q \longrightarrow q+\dfrac{d\phi}{dx} \, .
\label{eq:trasf_q}
\end{equation}
\end{subequations}
where $V_\mathrm{{ext}}(x)$ is the external potential associated with the spatial constraints in which the Fermi superfluid is embedded, as it is done in the original LPDA approach \cite{Simonucci-2014}. 
However, this simple procedure fails for the mLPDA approach, since it yields unphysical singularities in the diagonal element $\Gamma_{11}(Q,\mathbf{q})$ of the particle-particle ladder \cite{Pisani-2023},
to the extent that the gapless condition at $Q=0$ is not satisfied locally at each spatial point.
In practice, what one needs to satisfy locally the gapless condition at $Q=0$ is to replace the external potential $V_\mathrm{{ext}}(x)$ by a suitable \emph{effective\/} potential $V_{\mathrm{eff}}(x)$, such that
\begin{equation}
\mu\longrightarrow\mu-V_{\mathrm{eff}}(x)
\end{equation}
in the place of Eq.~(\ref{eq:locmu}), where $V_{\mathrm{eff}}(x)$ is constructed to ensure the gapless condition to be locally satisfied \cite{Pisani-2023} 
(explicit examples of $V_{\mathrm{eff}}(x)$ when $V_{ext}(x)$ has a Gaussian profile are given in Ref.~\cite{Pisani-2023}).
As a result, the local particle and current densities become:
\begin{subequations}
\begin{equation}
n(x)=\dfrac{2}{\beta}\sum_ne^{i\omega_n\eta}\int\!\!\dfrac{\text{d}\mathbf{k}}{(2\pi)^3}\mathcal{G}_{11}^{\mathrm{pf}}(\mathbf{k},\omega_n;\mathbf{q}|x)
\end{equation}
\begin{equation}\label{eq:current_mLPDA}
\begin{split}
j(x)=\dfrac{1}{m}\left(q+\dfrac{\text{d}\phi(x)}{\text{d}x}\right)n(x)\\
+\dfrac{2}{\beta}\sum_ne^{i\omega_n\eta}\int\!\!\dfrac{\text{d}\mathbf{k}}{(2\pi)^3}\dfrac{k_{x}}{m}\mathcal{G}_{11}^{\mathrm{pf}}(\mathbf{k},\omega_n;\mathbf{q}|x).
\end{split}
\end{equation}
\end{subequations}
These expressions correctly reduce to Eqs.~\eqref{eq:density_LPDA} and \eqref{eq:current_LPDA} when the effect of pairing fluctuations is neglected, since in this case the expression \eqref{eq:Green_pf} for $\mathcal{G}_{11}^{\mathrm{pf}}$ recovers
the expression \eqref{eq:Green_mf} for $\mathcal{G}_{11}^{\mathrm{mf}}$.

Finally, when considering the Josephson effect within the mLPDA approach \cite{Piselli-2023}, we solve the differential equation \eqref{eq:LPDA} subject to the condition $j(x)=J$ (where $J$ is a constant) for the current conservation, in analogy to what was done in Ref.~\cite{PSS-2020} within the LPDA approach in the absence of pairing fluctuations, but now $j(x)$ is given by the expression \eqref{eq:current_mLPDA} which includes pairing fluctuations.

\vspace{-0.3cm}
\section{Numerical results \\ for a Gaussian barrier} 
\label{sec:Numerical results}

In this Section, we present numerical results obtained by solving the LPDA equation and its mLPDA variant, in the presence of a barrier with a Gaussian profile along the direction $x$ of the supercurrent while the superfluid system is homogeneous
in the orthogonal ($y$ and $z$ directions).
Accordingly, the barrier has the form
\begin{equation}
V_{\mathrm{ext}}(x) = V_0\exp\left(\dfrac{x^2}{2\sigma_L^2}\right)
\label{eq:barrier}
\end{equation}
with $V_0=0.1E_F$ and $\sigma_L=2.5k_F^{-1}$ for definiteness.
The results presented in this Section are meant to be preliminary to those presented later in Sec.~\ref{sec:experiments}, where a more realistic spatial environment will have to be considered to account for the experiments on the Josephson effects
with ultra-cold Fermi gases reported in Refs.~\cite{Kwon-2020,DelPace-2021}.
In addition, we limit here to considering the coupling $(k_Fa_F)^{-1}=0$ and the temperature $T/T_c=0.15$.

\begin{figure}[t]
\begin{center}
\includegraphics[width=7.5cm,angle=0]{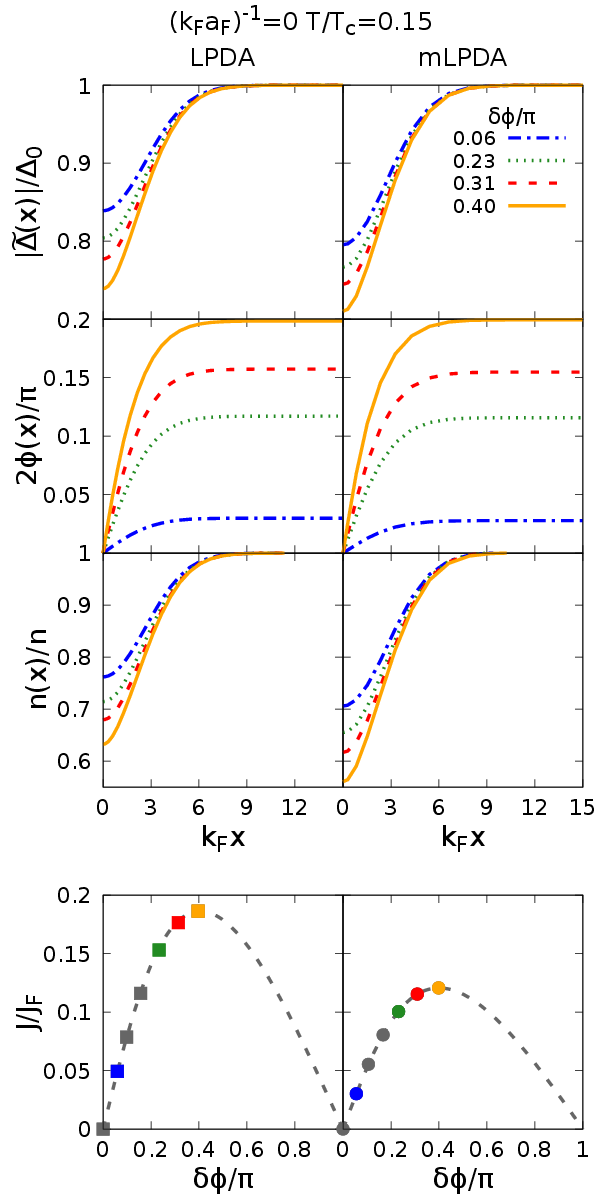}
\caption{Upper panels: Profiles of the magnitude $\tilde{|\Delta}(x)|$ and phase $2 \phi(x)$ of the gap parameter and of the density $n(x)$.
              Lower panels: Josephson characteristics $J(\delta\phi)$ vs the asymptotic phase difference $\delta \phi$ across the barrier (\ref{eq:barrier}). 
              These results are obtained by solving the LPDA equation (filled squares - left panels) and the mLPDA equation (filled dots - right panels) at unitarity and $T/T_{c}=0.15$,
              where the critical temperature $T_{c}$ is obtained within either the LPDA or mLPDA approaches.
             Here, $\Delta_{0}$ and $n$ are the bulk values of the (magnitude of the) gap parameter and density, respectively, and $J_{F}=k_{F}n/m$.
             [Reproduced from Fig.~3 of Ref.~\cite{Pisani-2023}.]}
\label{Figure-1}
\end{center} 
\end{figure}  

Figure~\ref{Figure-1} shows a number of profiles of the magnitude $|\tilde{\Delta}(x)|$ and phase $2\phi(x)$ of the gap parameter and of the density $n(x)$, for several values of the asymptotic phase difference $\delta\phi$ accumulated across the barrier. 
Note how most of the variation of $2\phi(x)$ is accommodated in the minimum of the density.
Here, each spatial profile corresponds to a point in the $J$ vs $\delta\phi$ characteristics that are also reported in the figure (as made evident by the matched colors for profiles and points).

Note from Fig.~\ref{Figure-1} that, for both LPDA and mLPDA approaches, the spatial profiles of $|\tilde{\Delta}(x)|$ vary over the width $\sigma_L$ of the barrier, which is larger than the Cooper pair size $\xi_{\mathrm{pair}}$ that at unitarity is of the order of $k_{F}$ (cf., e.g., Fig.~7 of Ref.~\cite{SPS-2010}).
This penetration of the gap parameter under the barrier is a characteristic fingerprint of the ``proximity effect'' \cite{Deutscher-1969,Piselli-2018}, which can be accounted for only when the contribution of the kinetic energy is accurately taken into account 
by the theoretical approach.
This is what occurs in both the LPDA and mLPDA approaches which inherit the correct treatment of the kinetic energy directly from the BdG equations \cite{BdG-1966}, such that by imposing appropriate boundary conditions one can take into account any shape (in particular, the height and width) of the barrier potential \cite{SPS-2010}.
We shall return to this important issue in Sec.~\ref{sec:experiments} when comparing with the available experimental data in ultra-cold fermi gases, for which the Cooper pair size is comparable or even smaller than the barrier width.

The LPDA and mLPDA approaches, however, give different values for $|\tilde{\Delta}(x=0)|/\Delta_0$ deep inside the barrier.
This can be seen from Fig~\ref{Figure-1},  where the normalized profiles $|\tilde{\Delta}(x)|/\Delta_0$ and $n(x)/n$ obtained within the mLPDA approach reach lower values in the vicinity of the barrier with respect to those obtained within the LPDA approach. 
This feature results, in turn, in smaller values of the critical current $J_c$ in the mLPDA with respect to the LPDA approach, as it is evident from the bottom panels in Fig~\ref{Figure-1}.
In Ref.~\cite{Pisani-2023} it was shown that this feature of the critical current $J_c$ at low temperature is not specific of unitarity but holds across the whole BCS-BEC crossover.

In addition, in Ref.~\cite{Pisani-2023} it was  shown that for increasing temperature the situation gets reversed, in the sense that, at a given coupling-dependent temperature measured with respect of $T_{c}$ (which is itself calculated for each coupling 
either within the LPDA or the mLPDA approach), the critical current $J_c$ obtained by the LPDA approach overcomes that obtained by the mLPDA approach.
It was also found in Ref.~\cite{Pisani-2023} that, while within the LPDA approach the overall shape of the $J_c$ vs $T/T_{c}$ curves remains the same for varying coupling, within the mLPDA approach this shape evolves from a convex to a concave behavior when passing from the BCS to the BEC regime, in such a way that at unitarity it appears essentially linear.

All these features will conspire against each other when considering more complex spatial geometries, like those utilized in the experiments with ultra-cold Fermi gases as discussed in the next Section.

\section{Comparison with experimental data for ultra-cold Fermi gases}
\label{sec:experiments}

A good benchmark of our numerical calculations for the Josephson critical current in complex spatial geometries are the recent experimental measurements in ultra-cold atomic Fermi gases,
both at low temperature as a function of coupling \cite{Kwon-2020} and at unitarity as a function of temperature \cite{DelPace-2021}. 
In practice, this represents an important test on the mLPDA approach of Ref.~\cite{Pisani-2023}, to the extent that the explanation of experimental data is one of the cornerstones of what theory should be about.

In both experiments, $^6$Li atoms are confined by a potential of the form
\begin{equation}\label{eq:trapping_potential}
V_{\mathrm{trap}}(x,y,z) = \dfrac{1}{2}m\left(\omega_x^2x^2+\omega_y^2y^2+\omega_z^2z^2\right) \, ,
\end{equation}
where $m$ is the atomic mass and $\omega_{x,y,z}=2\pi(12,165,140)$ Hz in Ref.~\cite{Kwon-2020} and $\omega_{x,y,z}=2\pi(17,300,290)$Hz in Ref.~\cite{DelPace-2021}. 
To enhance the stability of the system, the atomic cloud is further shaped by two walls raised at $\pm x_w$ (with $k_F^tx_w=187.1$ in Ref.~\cite{Kwon-2020} and $k_F^tx_w=253.3$ in Ref.~\cite{DelPace-2021}).
Here, $k_F^t=\sqrt{2mE_F^t}$ is the trap Fermi wave vector associated with the trap Fermi energy $E_F^t=\omega_0(3N)^{1/3}$, where $\omega_0$ is the geometrical mean of the trapping frequencies and $N$ the number of $^6$Li atoms 
(with $N=2.6\times10^5$ in Ref.~\cite{Kwon-2020} and $N=3.0\times10^5$ in Ref.~\cite{DelPace-2021}).
The raising of walls decreases the number of the trapped atoms from the original value $N$ to a new value $N_w$, such that $N_w=(1.0\div1.4)\times10^5$ in Ref.~\cite{Kwon-2020} and $N_w=1.6\times10^5$ in Ref.~\cite{DelPace-2021}.
A barrier is eventually raised at the center of the atomic cloud, whose form is
\begin{equation}
V_{\mathrm{barrier}}(x,y,z) = V_0(z)\exp\left(-2\dfrac{x^2}{w(z)^2}\right) \, ,
\label{eq:trap_barrier}
\end{equation}
where
\begin{subequations}
\begin{equation}
V_0(z)=V_0\Bigg/\sqrt{1+\left(\dfrac{z}{z_R}\right)^2} \, ,
\label{eq:V-0-z}
\end{equation}
\begin{equation}
w(z)=w_0\sqrt{1+\left(\dfrac{z}{z_R}\right)^2} \, ,
\label{eq:w-z}
\end{equation}
\end{subequations}
with $V_0/E_F=(0.38,0.455,0.52)$ and $k_F^tz_R=14.24$ in Ref.~\cite{Kwon-2020} and $V_0/E_F^t=0.411$ and $k_F^tz_R=26.07$ in Ref.~\cite{DelPace-2021}. 
In this case, the overall external potential is the sum of three contributions, namely, 
\begin{eqnarray}
& V_{\mathrm{ext}}(x,y,z) & \, = V_{\mathrm{trap}}(x,y,z) 
\nonumber \\
& + & \hspace{-0.6cm} 1.2 \times 10^{3} E_{F}^{t} \, \left\{ \theta[(k_{F}^{t}(x-x_{\mathrm{w}})] + \theta[-(k_{F}^{t}(x+x_{\mathrm{w}})] \right\} 
\nonumber \\
& + & V_{\mathrm{barrier}}(x,y,z) \, .
\label{V-ext} 
\end{eqnarray}
Additional details can be found in Ref.~\cite{Piselli-2023}.

\begin{figure}[t]
\begin{center}
\includegraphics[width=8.0cm,angle=0]{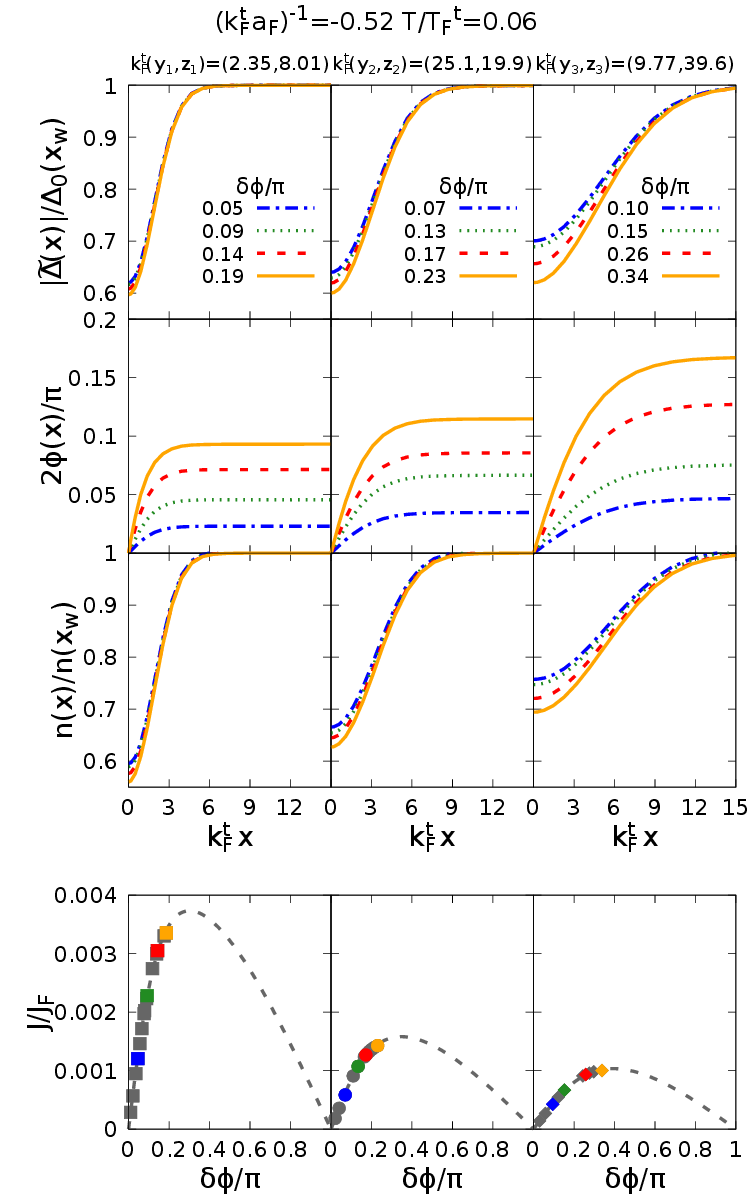}
\caption{Upper panels: Profiles of the magnitude $|\tilde{\Delta}(x,y_{i},z_{i})|$ and phase $|\tilde{\Delta}(x,y_{i},z_{i})|$ of the gap parameter and of the density $n(x,y_{i},z_{i})$, for each of the three chosen filaments with $i = (1,2,3)$.
              Lower panels: Corresponding Josephson characteristics $J(\delta\phi)$ (in units of $J_{F} = k_{F} n/m$ with $n$ the asymptotic density at the edges $\pm x_{w}$ of the cylinder) for the current vs the asymptotic phase difference 
              $\delta \phi$ across the cylinder, where the dashed lines represent a fitting procedure to the filled dots (with the same notation as in Fig.~\ref{Figure-1}).
              All these results are obtained by solving the mLPDA equation for $(k_{F}^{\mathrm{t}} a_{F})^{-1} = -0.52$ and $T/T_{F}^{\mathrm{t}} = 0.06$.}
\label{Figure-2}
\end{center} 
\end{figure}  

With this nontrivial geometry to deal with, numerical solution of the mLPDA equation should, in principle, be supplemented by imposing the continuity equation $\nabla\cdot\mathbf{j}(\mathbf{r}) = 0$ in 3D for the local particle current $\mathbf{j}(\mathbf{r})$.
Fortunately this is not the case, and in the following we can safely assume that $j_y=j_z=0$ and solve the mLPDA equation with the simpler one-dimensional condition $d \, j_{x}(x) / dx = 0$.
This is because from an independent numerical calculation performed in the BEC limit in terms of the time-dependent Gross-Pitaevskii equation \cite{Simonucci-unpublished}, which directly simulates the experimental set up where the barrier moves with a steady velocity through the superfluid at rest, the current lines are found to remain straight as long as the velocity of the barrier is below a critical value.

Further neglect of the trapping frequency $\omega_x$ allows us to assimilate the truncated ellipsoid to a cylinder and associate to each pair of coordinates $(y,z)$ a uniform current density. 
Accordingly, we divide the cylinder into a bundle of (at most 441) tubular filaments and solve the mLPDA equation for each of these filaments, similarly to what we did in Sec.~\ref{sec:Numerical results}. 
Finally, by adding up the contributions from all filaments, we draw the ``global'' Josephson characteristic $I(\delta \phi)$ for the cylinder as a whole and identify the critical current $I_{c}$ from its maximum.

As an example of this procedure, we have selected three tubular filaments with $i = (1,2,3)$ identified by their transverse coordinates $k_{F}^{\mathrm{t}} (y_{1},z_{1}) = (2.35,8.01)$, $k_{F}^{\mathrm{t}} (y_{2},z_{2}) = (25.1,19.9)$, and $k_{F}^{\mathrm{t}} (y_{3},z_{3})= (9.77,39.6)$. 
For each of these filaments, Fig.~\ref{Figure-2} shows the profiles of the magnitude $|\tilde{\Delta}(x,y_{i},z_{i})|$ and phase $2 \phi(x)$ of the gap parameter and of the density $n(x,y_{i},z_{i})$ along the $x$ direction, for several values of the asymptotic phase difference $\delta\phi$ accumulated across the two sides of the cylinder, in analogy to what is done in Fig.~\ref{Figure-1} for the simpler one-dimensional Gaussian barrier (although in Fig.~\ref{Figure-2} only results obtained by the mLPDA approach are explicitly reported).
For later convenience, the numerical calculations shown in Fig.~\ref{Figure-2} are performed for coupling $(k_{F}^{\mathrm{t}} a_{F})^{-1} = -0.52$ and temperature $T/T_{F}^{\mathrm{t}} = 0.06$,
where $T_{F}^{\mathrm{t}}$ is the trap Fermi temperature.
In addition, the lower panels of Fig.~\ref{Figure-2} show the Josephson characteristics corresponding to each filament.
Note that, the closer the filament lies to the axis of the cylinder, the more the Josephson characteristics deviates from the simple $\sin (\delta \phi)$ behavior obtained for a tunneling barrier \cite{SPS-2010}, since the local chemical potential $\mu - V_{\mathrm{ext}}(\mathbf{r})$ correspondingly tends to exceed the maximum height $V_{0}(z=0)$ of the barrier.

\begin{figure}
\begin{center}
\includegraphics[width=7.2cm,angle=0]{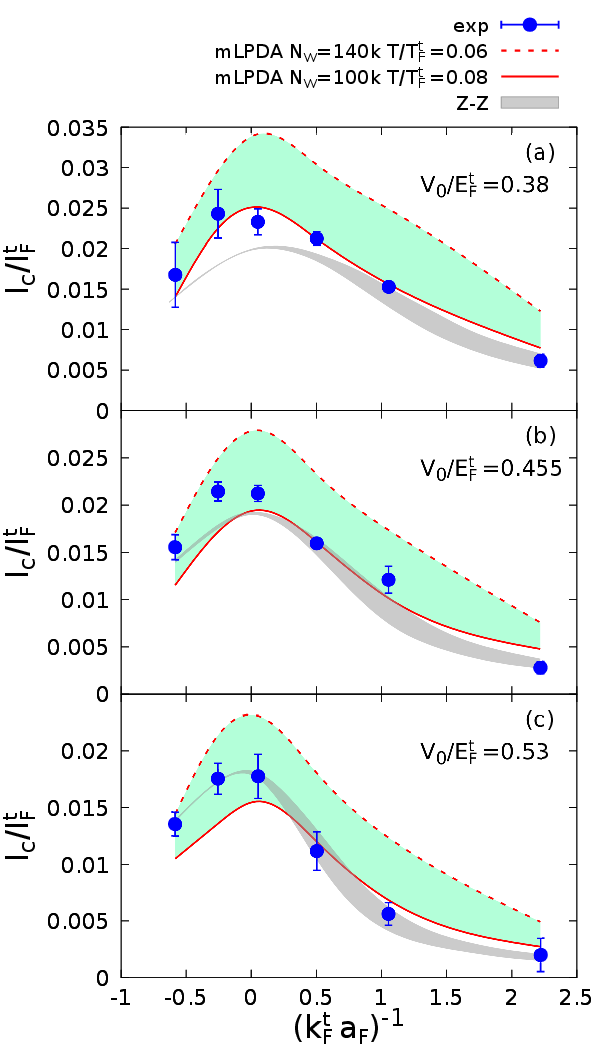}
\caption{Critical current $I_{c}$ (in units of $I_{F}^{t}$) vs the trap coupling $(k_{F}^{t} a_F)^{-1}$, for three barriers with the same width and different heights. 
              The experimental data from Ref.~\cite{Kwon-2020} (dots with error bars) are compared with the theoretical results obtained by solving the mLPDA equation (green stripes bound by an upper (dashed) and lower (full) red curve).
              These curves correspond to the experimental uncertainties on temperature and $N_{w}$ reported on top of the figure (plus a $5 \%$ uncertainty on $w_{0}$ of Eq.~(\ref{eq:w-z})).
              Reported are also the theoretical (Z-Z) results obtained by the approach of Ref.~\cite{ZZ-2019} (grey stripes), where the $5 \%$ uncertainty on $w_{0}$ is only considered.
              [Adapted from Fig.~3 of Ref.~\cite{Piselli-2023}.]}
\label{Figure-3}
\end{center}
\end{figure}

Once the contributions $j(\delta \phi)$ from each filament are added together, the value of the critical current $I_{c}$ can be extracted from the maximum of the characteristics $I(\delta \phi)$ of the total current 
(and conveniently normalized in terms of the quantity $I_{F}^{t} = k_{F}^{t} \int \!\! dy dz \, n(x_{\mathrm{w}},y,z)/m$). 
Figure~\ref{Figure-3} shows the results of this procedure implemented for several couplings within the mLPDA approach for  three barriers of the form (\ref{eq:trap_barrier}) considered experimentally in Ref.~\cite{Kwon-2020}, 
together with the experimental data reported in the same reference (blue filled circles with error bars).
Specifically, the theoretical mLPDA results correspond to the green stripes, which are bound by an upper (dashed) and lower (full) red curve associated with different temperatures and different values of the total number of atoms $N_{w}$ contained
in the atomic cloud after raising the walls (which correspond, in turn, to the experimental uncertainties on temperature and $N_{w}$ \cite{Kwon-2020,Roati_private}).
A further $5 \%$ uncertainty in the value of $w_{0}$ of Eq.~(\ref{eq:w-z}) is also taken into account.
Note how the mLPDA results provide a remarkably good quantitative agreement with the experimental data over wide physical conditions.

Reported in Fig.~\ref{Figure-3} are also the theoretical results (grey stripes) obtained at zero temperature by the method of Ref.~\cite{ZZ-2019}, that were as well reported in Fig.~3 of Ref.~\cite{Kwon-2020}, 
which take into account a $5 \%$ uncertainty on $w_{0}$ but fix the number of atoms to the value $N$ before the raising of the walls (which is about twice as much the value of $N_{w}$ eventually contained in the 
experimental cloud \cite{Kwon-2020,Roati_private}, although this difference does not affect the value of the local chemical potential).
Note how in this case the comparison with the experimental data becomes worse as the height $V_{0}$ of the barrier decreases, which is consistent with the fact that the approach of Ref.~\cite{ZZ-2019} relies on a tunneling approximation
which holds for a strong enough barrier when the details of the profile of the gap parameter inside the barrier are not relevant.
The mLPDA approach, on the other hand, provides a full account of these details.

\begin{figure}
\begin{center}
\includegraphics[width=9.0cm,angle=0]{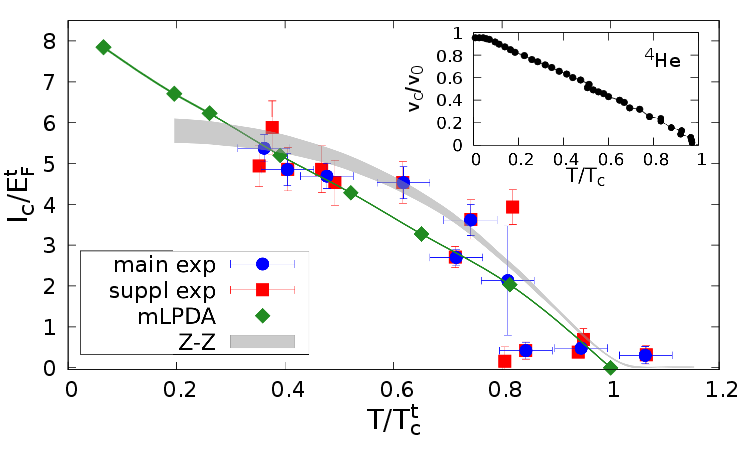}
\caption{Temperature dependence (in units of the trap critical temperature $T_{c}^{t}$) of the critical current $I_{c}$ (in units of $E_{F}^{t}$) at unitarity.
              The experimental data, from the main text (blue dots with error bars) and the supplemental material (red squares with error bars) of Ref.~\cite{DelPace-2021}, are compared with the theoretical results obtained 
              by the mLPDA approach of Ref.~\cite{Pisani-2023} with $N_{\mathrm{w}} = 160k$ (green diamonds) and by the approach (Z-Z) of Ref.~\cite{ZZ-2019} (grey stripe), once extended to deal with finite temperatures 
              as mentioned in Ref.~\cite{DelPace-2021}.
              The inset shows the critical velocity of superfluid $^{4}\mathrm{He}$ (normalized to its zero-temperature value), as extracted from Fig.~12 of Ref.~\cite{Varoquaux-2015}. 
              [Adapted from Fig.~4 of Ref.~\cite{Piselli-2023}.]}
\label{Figure-4}
\end{center}
\end{figure}

But there is also another issue for which the predictions of the mLPDA approach of Ref.~\cite{Pisani-2023} differ from those of the approach of Ref.~\cite{ZZ-2019}.
It concerns the temperature dependence of the critical current $I_{c}$ at unitarity, once compared with the experimental data reported in Ref.~\cite{DelPace-2021}.
This comparison is shown in Fig.~\ref{Figure-4} for a barrier with $V_0/E_F^t=0.411$ \cite{DelPace-2021}, which is close to the smallest value considered in Fig.~\ref{Figure-3}.
Note from this figure how the mLPDA approach yields a \emph{linear\/} behavior for $I_{c}$ vs $T$, while in the approach of Ref.~\cite{ZZ-2019} a concave behavior is obtained.
The latter originates from the temperature dependence of the condensate density $n_{0}$, to which $I_{c}$ is directly related in the approach of Ref.~\cite{ZZ-2019}.
Within the mLPDA approach, on the other hand, it was found in Ref.~\cite{Pisani-2023} that, for the simple barrier here also considered in Sec.~\ref{sec:Numerical results}, the temperature dependence of the critical current changes from a convex to a concave behavior from the BCS to the BEC regime, passing through an intermediate linear behavior at unitarity.
This finding is consistent with the linear behavior joining the green diamonds in Fig.~\ref{Figure-4}, which is obtained by the mLPDA approach once the experimental geometrical constraints corresponding to the external potential (\ref{V-ext})  are taken into account.

For the sake of comparison, the inset of Fig.~\ref{Figure-4} shows that a linear temperature behavior is shared by the critical velocity of superfluid $^{4}\mathrm{He}$, as reproduced from Fig.~12 of Ref.~\cite{Varoquaux-2015}.
That $^{4}\mathrm{He}$ and the unitary Fermi gas share similar behaviors under several circumstances has been already emphasized in the literature, ranging from the superfluid fraction \cite{Sidorenkov-2013}
to the sound propagation \cite{Kuhn-2020}, and here noted for the temperature dependence of the critical current.

\section{A potential relation between \\ the Josephson effect and \\ the superfluid density}
\label{sec:Josephson_vs_superf-dens}

A key property of a superfluid is the superfluid density $\rho_{s}$ \cite{Landau-1941}.
A \emph{direct\/} (and experimentally measurable) connection between $\rho_{s}$ and the Josephson effect was recently demonstrated in Ref.~\cite{Modugno-2023}, where it was however pointed out that this connection is applicable only to a supersolid and not to a conventional superfluid.
In this Section, we highlight the existence of an \emph{indirect\/} connection between $\rho_{s}$ and the Josephson effect also for conventional superfluids of interest in the present article, suggesting at the same time an explanation of the fact that in Fig.~\ref{Figure-4}
the predictions of the mLPDA approach of Ref.~\cite{Pisani-2023} differ from those of the approach of Ref.~\cite{ZZ-2019}.

To this end, we recall that in Ref.~\cite{ZZ-2019} a ``tunneling approach'' was adopted, whereby the contribution to the current from atoms excited out of the condensate was neglected.
On the other hand, in the mLPDA approach of Ref.~\cite{Pisani-2023} it is the superfluid density $\rho_{s}$ (and not the condensate density $n_{0}$) to determine the dynamical properties of the superfluid flow, 
in such a way that both pairs in the condensate and pairs excited out of the condensate by quantum and thermal fluctuations contribute to the superfluid flow. 
This is related to the fact that, both in the original LPDA approach of Ref.~\cite{Simonucci-2014} and its improved mLPDA version of Ref.~\cite{Pisani-2023}, the expression of the supercurrent is consistent with that of a two-fluid model.

In this context, a key argument was pointed out in Ref.~\cite{Pisani-2023-II} (cf. Sec.~III-D therein), where the Josephson critical current flowing through a given barrier was interpreted as an intrinsic critical current inside that barrier, namely, 
the current corresponding to a bulk superfluid but subject to the ``local" thermodynamic conditions at \emph{the center\/}  ($x=0$) \emph{of the barrier\/}.
Here, the value $\rho_s(x=0)$ can be obtained within a Local Density Approximation (LDA) perspective from the local condition $j = \left( q + \frac{d \phi(x)}{dx}|_{x=0} \right) \rho_s(x=0)$ for the supercurrent, 
where the value of $\frac{d \phi(x)}{dx}|_{x=0}$ is instead provided by the mLPDA approach.
To contrast the results of Ref.~\cite{Pisani-2023} with those Ref.~\cite{ZZ-2019} the local value $n_0(x=0)$ of the condensate density is also required.
This can be obtained from the results of Ref.~\cite{Pisani-2022} for a homogeneous superfluid, in which the local values of the gap parameter and chemical potential associated with the same barrier are now utilized, again within a LDA perspective.

In particular, for the Gaussian barrier considered in Sec.~\ref{sec:Numerical results} (of height $V_{0}/E_{F}=0.1$ and width $\sigma_{L} = 2.5 k_{F}^{-1}$), at unitarity and for $T/T_{c} = 0.08$ we obtain
$n_0(x=0)=25\%$ (with respect to half the bulk density $n$) and $\rho_s(x=0)=70\%$ (with respect to the full bulk density).
By increasing the barrier height to $V_{0}/E_{F} = 0.3$, we obtain instead $n_0(x=0)=10\%$ and $\rho_s(x=0)=n(x=0)=25\%$.
(In both cases, we find that $\rho_s(x=0)$ coincides with the local value $n(x=0)$ of the density as expected.)
These numerical results signal that $n_0(x=0)$ and $\rho_s(x=0)$ tend to approach each other percentage wise as the barrier height is increased.
Extrapolating this trend to the limit when the barrier height $V_{0}$ approaches $E_{F}$, we are led to the conclusion that in this ``tunneling'' limit the physics of the Josephson effect can be directly related to the condensate density. 
This is in line with what was assumed by the tunneling approach of Ref.~\cite{ZZ-2019}, although that approach was then applied also to barrier heights and widths for which it should not be expected to hold.

Further support to our argument, about the relevance for the Josephson effect of the out-of-condensate density (and not only of $n_{0}$), comes from $^{4}\mathrm{He}$ (cf. the inset of Fig.~\ref{Figure-4}) for which $n_0$ is only $10\%$ and most of the superfluid flow is carried by
excited quasiparticles.
Finally, this picture is also in line with the measurements of the conductance made in Ref.~\cite{DelPace-2021}, where an anomalous increase of the conductance below $T_{c}$ was ascribed to bosonic collective excitations, 
which propagate through the quantum as well as thermal depletions of the condensate.

\section{Concluding remarks and perspectives}
\label{sec:conclusions}

In this work, we have reviewed the mLPDA approach developed in Ref.~\cite{Pisani-2023}, whereby pairing fluctuations beyond mean field are included in the presence of non-trivial geometries to account for the spatial profile of the gap parameter
under a wide range of physical circumstances.
These include specifically the BCS-BEC crossover realized with ultra-cold Fermi gases.
In particular, the method has been applied in Ref.~\cite{Piselli-2023} to account for recent experiments on the Josephson effect realized with these gases.
Besides validating the mLPDA approach by a favorable comparison with these experiments, here we have widened the discussion on what physical knowledge one can get from this comparison, also by contrasting with the results of competing theoretical approaches.


	

\end{document}